# Computational prediction of RNA tertiary structures using machine learning methods

Bin Huang (黄斌), Yuanyang Du (杜渊洋), Shuai Zhang (张帅), Wenfei Li (李文飞), Jun Wang (王骏), Jian Zhang (张建)†

*National Laboratory of Solid State Microstructures, School of Physics, Collaborative Innovation Center of Advanced Microstructures, Nanjing University, Nanjing 210093*
*Institute for Brain Sciences, Kuang Yaming Honors School, Nanjing University, Nanjing 210093*

**Abstract**

RNAs play crucial and versatile roles in biological processes. Computational prediction approaches can help to understand RNA structures and their stabilizing factors, thus providing information on their functions, and facilitating the design of new RNAs. Machine learning (ML) techniques have made tremendous progress in many fields in the past few years. Although their usage in protein-related fields has a long history, the use of ML methods in predicting RNA tertiary structures is new and rare. Here, we review the recent advances of using ML methods on RNA structure predictions and discuss the advantages and limitation, the difficulties and potentials of these approaches when applied in the field.

*Project supported by the National Natural Science Foundation of China (Grant No. 11774158, 11974173, 11774157, and 11934008).
†corresponding authors: jzhang@nju.edu.cn

**Introduction**

RNAs are macromolecules of crucial and versatile biological functions.[1-5] To fully understand their functions, knowledge of the three-dimensional (3D) structures is essential. Since experimental approaches to determinate RNA 3D structures are difficult and expensive, many computational approaches have been developed to this purpose. To date, although template-based and homology-modeling methods could achieve high accuracies, *de* novo predictions still depends on the size and complexity of the RNA, and further improvement in predicting non-canonical interactions are required, according to the recent RNA-Puzzles round III.[6] For a comprehensive study of the recent work, we refer readers to the relevant literature.[7-11]

Recently, machine learning (ML) techniques, particularly deep learning based on multiplayer neural networks, have achieved great success in characterizing, classifying, and/or generating complex data in a broad range of fields, from image classification, disease diagnosis, solving biological problems, playing chess, or games to even quantum physics.[12-17] It is exciting to investigate whether these techniques could assist in RNA tertiary structure predictions.

In this short review, we summarize the recent advances and applications of ML techniques in solving the RNA tertiary structure prediction, including our current work. We wish this could develop new ideas and draw novel prediction algorithms for predicting. While there are many excellent works of using ML techniques on predicting RNA secondary structures,[18-20] here we focus on the significance of RNA tertiary structure predictions.

**Multilayer perceptron architecture for tertiary structure scores**

An important step for structure prediction is to evaluate the generated structure of candidates. To this purpose, many scoring functions have been developed, and most of them are based on the inverse Boltzmann equation. However, these methods need to carefully choose the proper function forms and reference states, as is not an easy task.[8,21-22] Furthermore, the functions are usually composed of only pair-wise terms. While incorporating many-body interactions is possible, on the contrary determining the relevant parameters is practically hindered by the lack of sufficient experimental data.

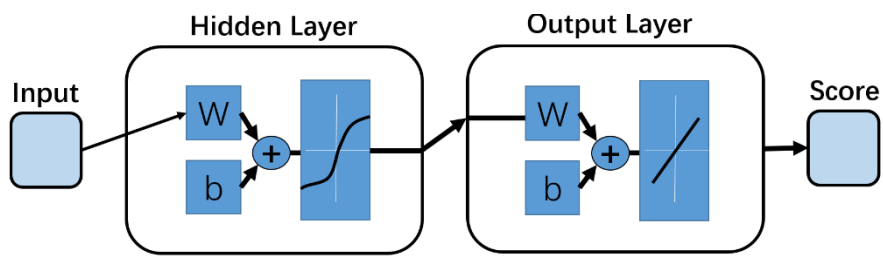

Figure 1. The architecture of the multilayer perceptron used in the work.[23] It contains a single hidden layer. The inputs are structural features, and the output is a score that indicates the quality of the structural candidates.

We built a scoring system using multilayer perceptrons to score the RNA tertiary structure candidates. The system is significantly different from the traditional scoring functions.[23] The theoretical basis of this approach is based on the universal approximation theorem, which states that a feed-forward network with a single hidden layer can approximate a wide variety of continuous functions when given appropriate parameters.[12] We built two feed-forward multilayer perceptrons, labeled as Net-1 and Net-2, respectively. They are different in the input features. Specifically, Net-1 accepts as inputs the features calculated at the coarse-grained (CG) structural level, while Net-2 accepts as inputs the all-atom structural features. The output of either network is a score, which is the similarity of the input structure to the native one. In this work, the similarity is measured with the RMSD of the input structure after optimal superimposition with the corresponding experimental one. The choice of input features allows great flexibility. In principle, the features can be anything of interest. Here we cautiously chose the probabilities of observing base couples of different types and the probabilities of backbone atom pairs (at the CG level) at different distance bins as inputs for Net-1, and the probabilities of observing atom pairs (at the all-atom level) of different types at different distance bins for Net-2. Taking a standard A-helix of 50-nucleotides in length as an example, a simple statistic gives a total number of atom pairs of 908. Among them 48 P-C4' pairs are within the distance of 0.3–0.4 nm, then the probability of observing the type P-C4' at the third distance bin will be set to 48/908, if the bin size is set to 0.1nm. Moreover, the number of nucleotides of the candidate structure and its radius of gyration along three principal axes are also inputted to the networks, in order to feed the size and shape information to the neural network.

The setups of the multilayer perceptron are described as follows. As shown in Figure 1, both perceptrons contain a single hidden layer, which contains 30 nodes for Net-1 and 10 nodes for Net-2, respectively. These numbers were obtained by optimizing the scoring performance. The activation function for the neurons in the hidden layer is the hyperbolic tangent function, and that for the neurons in the output layer is the linear function. The loss function is the weighted mean squared error between the predicted score and the RMSD with regards to the native structure. The weight is proportional to the exponential of the negative values of "R", where R is the RMSD of the sample, in order to encourage the contribution of the close-to-native samples to the loss. Furthermore, the weight is inversely proportional to the number of samples in the corresponding RMSD bin, which helps to increase the contribution of rare samples. The networks were optimized by the Gradient Backpropagation algorithm for reducing the difference between the predicted scores and the true values.

The dataset contains 462 RNAs, each associated with 300 decoys. The decoys were generated with molecular dynamics simulations (GROMACS v4.5) with the temperature gradually increased from 300K to 600K. For each RNA, 300 structures were randomly taken from the trajectories in a way that their distribution was uniform in the RMSD range [0, $RMSD_{max}$], where the upper bound is dependent on the RNA length. In total we obtained 138,600 structures. The dataset was randomly split into the training, validation, and testing dataset with the ratio 322:70:70. The multilayer perceptrons were trained, validated, and tested with these three datasets, respectively.

The performance of the networks was encouraging. For the test dataset containing 70 RNAs and the associated decoys, the correlations between the predicted and the true values were generally good and showed a funnel-like shape.[23] The performance was also compared with RASP, [25] which is a state-of-the-art all-atom knowledge-based potential for assessing RNA 3D structures. RASP explicitly accounts for base pairing, base stackings and non-canonical interactions that are highly abundant in RNA structures. RASP was showed to be competitive when compared to NAST, ROSETTA, and AMBER force fields. The comparison of our models with RASP showed that, the enrichment score, which measures the degree of overlap between the best-scored structures (10%) and the most native-like structures (also 10%),[24] was 4.6 for Net-1 and 5.3 for Net-2 on average, while was 4.4 for RASP for the same test set. As for the ability to select native structures out of decoys, Net-1 and Net-2 ensured that the native ones were among the top-10 scored structures for 60 out of 70 RNAs, and 52 out of 70 RNAs, respectively. In contrast, RASP gave 31 out of 70 RNAs with the same criterion.

The multilayer perceptron has many advantages in scoring structural candidates. First, it could accept any features as inputs, not just limited to the pair-wise features. Second, the weights in the networks can be trained in an end-to-end way. Third, it avoids to determinate the reference state, as it is known to be difficult in the traditional way of developing scoring functions.[21-22] However, the work described above is rather preliminary and it can be improved significantly. First, most inputs to the network are still pair-wise features, due to the lack of enough training data. More complex features such as many-body interactions will be possible with time, as more experimental structures become available. Second, the network architectures are rather simple. Presumably, deep networks may give better results, as it is generally believed in the image recognition field that deep networks usually have better performance.

**RNA tertiary structure scoring with convolutional neural networks**

RNA structure prediction community may borrow ideas from the latest image recognition field, where deep convolutional neural networks (CNNs) have become prevalent tools for various tasks of image processing. One remarkable characteristic of CNNs is their ability to extract complex patterns after proper training, in contrast to early times when people had to develop different operators manually in order to detect different patterns in images. In the RNA work described above, we had to determine relevant factors and invent input forms for these factors manually. Inspired by the advance in the image recognition field, we tried to simplify this process and let the neural network extract the most relevant features automatically.

We built a convolutional neural network to extract structural patterns from RNA tertiary structures and score those structures.[26] The CNN is a VGG-like network,[27] containing four convolutional layers with 8, 16, 32, and 64 convolutional filters, respectively, a max-pooling layer, and a fully connected layer, as shown in Figure 2. The input to the network is a 3D image of size $32 \times 32 \times 32$ voxels, obtained by gridding a cubic space of size $32 \times 32 \times 32$Å, with the interested nucleotide in the center. For a given RNA structure, the nucleotides are selected as the interested nucleotide one by one in the sequence order. The interested nucleotide is rotated to a specific direction and is put in the center of the cubic space, which is then subjected to a gridding procedure with a grid size of 1Å. After gridding, the cubic space is converted into a 3D image. The 3D image contains three channels, corresponding to the occupation number, mass, and charge in the grid, respectively. The output of the network is a score for the interested nucleotide, indicating the likeness of the local structure to the native structure. The score for the whole structure is the summary of the scores of all the nucleotides.

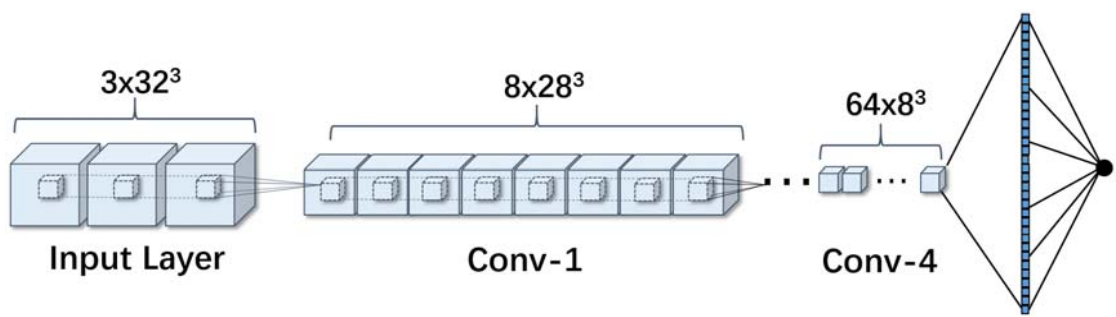

Figure 2. The architecture of the CNN network in this work.[26] Note that not all convolutional layers are shown due to space limitations. Each cube represents a 3D image. The input layer has three channels, similar to the RGB channels in 2D images. The output is a single score, indicating the likeness of the input structure to the native structure.

The training dataset was initially downloaded from the NDB website with the "RNA Only" and "Non-Redundant RNA structures" options, resulting in 619 RNAs. Then the RNAs with X-ray resolution > 3.5Å were removed, and these RNAs were further removed if they also appeared in the test dataset or belonged to the equivalent class as the RNAs in the test dataset. The test datasets were described subsequently. The final collection contains 414 RNAs. These RNAs were then subjected to high-temperature molecular dynamics (MD) simulations with Gromacs and Monte Carlo (MC) fragment assembly procedures with Rosetta[28-29] to generate decoys. The MD and MC procedures led to two training datasets, respectively; each containing 1.2 million samples. The MD based dataset was used to train a network for scoring RNA structures known to be close to the native structure, while the two datasets were combined to train another network for scoring general RNAs.

The performance of the CNN model was tested with three datasets. The first dataset contains 85 RNAs, each associated with 500 decoys generated with MODELLER.[25] The RMSDs of the decoys of most RNAs are less than 10 Å, while for some larger RNAs they can range from 0 to 13 Å. The second dataset contains 20 RNAs and 56500 decoys generated with REMD simulations or normal-model (NM) perturbations.[30] The RMSDs of the decoys generated with REMD range from 0 to 8 Å and those generated with NM perturbations range from 0 to 5 Å. The third dataset comes from the RNA-Puzzles competition from rounds I to III, containing 18 RNAs.[6,31-32] Each RNA has 12–70 predicted models, which were generated by the best existing RNA 3D structure predictors. The RMSDs with respect to the experimental structure range from 2 to 4 Å for some RNAs while range from 20 to 60 Å for several long RNAs with more than 100 nucleotides. In the testing procedure, each candidate structure in the testing set was fed to the CNN network and assigned a score. The structure with the best score was predicted as the native one, and then was compared with the experimental structure. The performance of the CNN network was compared with that of the other four popular scoring functions, including RASP,[25] Rosetta,[28-29] KB,[30] and 3dRNAscore.[33] The results are shown in Table 1.

Table 1. The performance of different scoring functions. In each cell, the first number is the number of RNAs that are correctly identified, and the second is the total RNAs in the dataset.[26] The bold number indicates the best one among the same dataset.

|  | 3dRNAscore | KB | RASP | Rosetta | CNN model |
|---|---|---|---|---|---|
| **dataset-I** | **84**/85 | 80/85 | 79/85 | 53/85 | 62/85 |
| **dataset-II** | 17/20 | **20**/20 | 12/20 | 12/20 | 19/20 |
| **dataset-III** | 5/18 | - | 1/18 | 4/18 | **13**/18 |

It can be seen that for the testing dataset-I, the CNN model ranks the fourth. For the dataset-II, it ranks the second but only slightly worse than the KB potential. While for the dataset-III, which is composed of structures from real competitions, the CNN model ranks significantly better than the others. Overall, the CNN model is competitive with the other state-of-the-art potentials. Considering that most decoys in the dataset-I and -II are obtained by perturbation of the native structure, the CNN model may need to be further trained with structures close to the native basin of attraction.

According to the performance of the CNN model, it is safe to infer that the model can extract relevant features out of data automatically. In order to understand what features the neural networks actually learn, we tentatively plotted the saliency maps by computing the gradient of the output score with respect to the input values.[26] The gradients indicate how sensitive the output is to the changes in the input, hence revealing what the important factors are for the structure. The analyzing results showed that the atoms with larger gradients in the local environment of the interested nucleotides generally correspond to base pairs or base stackings, consistent with the general physical knowledge for RNA structures.[26] However, this finding is dependent on visual inspection of experts and is thus rather preliminary. For a better understanding the NN, further collaborations with researches in computer science are needed to develop more advanced tools. In the near future, it is interesting to see if the neural networks could extract out of data new knowledge unknown before.

**Conformational sampling with machine learning approaches**

A powerful sampling engine that is able to efficiently and accurately generate 3D structural candidates is of paramount importance for a *de* novo tertiary structure predictor. Traditionally, 3D fragments, motifs, or small secondary elements databases are combined with MC, Las Vegas, game algorithm, or user-driven manipulation to generate decoys.[10] However, the discrete nature of the traditional methods imposes inherent problems on the coverage of the structural space and on the development of scoring functions.[34]

Frellsen et al. offered a different solution to the sampling problem by developing a probabilistic model of RNA tertiary structure that allows sampling in a continuous space.[34] The model, called BARNACLE, captures the marginal distribution of each of the seven torsional angles of one nucleotide and their dependence with a Dynamic Bayesian Network (DBN). Specifically, the model uses a slice for each angle and the hidden state of each slice is dependent on the present angle identifier and the hidden state of the previous slice along the chain. The model parameters are trained by maximum-likelihood estimation from experimental structures. The authors showed that the model captures the length distribution of helices, and is consistent with the RNA rotamer model. The authors also performed Markov chain MC simulations with the proposal distribution generated by the DBN, and found that the generated RNA decoys are similar to the experimental structures even guided with a simple energy function based solely on base pairs.

Wang et al. developed another probabilistic method for structure modeling and sampling. The method contains a conditional random fields model for tertiary structure and a tree-guided scheme for sampling, named TreeFolder.[35] Differently from BARNACLE, the above model estimates the probability of an RNA structure conditioned on the primary sequence and secondary structure. The authors showed that the model captures the structure-sequence relationship well and generates a much higher percentage of native-like decoys than the previous methods. Both BARNACLE and TreeFolder methods do not use fragments to build RNA structures, and their probabilistic nature allows an efficient as well as unbiased sampling of RNA conformations.

**Machine learning approaches in 3D modules identification**

On different line, computational efforts have been devoted to identifying 3D structural modules from single or multiple sequences. RNA modules are sets of recurrently observed non-Watson-Crick (WC) base pairs embedded between WC pairs. Their identifications are import for structure predictions since the non-WC interactions define RNA tertiary structures and constitute the main bottlenecks of current predictions.[36] In contrast to the traditional structure predictors that generate and evaluate different structures for a given sequence, these approaches scan and evaluate different sequences for a pre-defined structure or interaction pattern. RMDetect is a pioneer program in the field.[37] For a specific type of 3D structural module, RMDetect builds a Bayesian probabilistic network with the nodes representing individual bases occupying a defined structural position and the edges representing their dependence. Then, it scans the sequence database and calculates the compatible probability of the sequence with the 3D module by threading the sequence into the corresponding Bayesian network. RMDetect was initially designed to identify four types of modules, including G-bulge loop, kink-turn, C-loop, and tandem-GA loop, but it can be easily extended to other modules. The metaRNAmodule pipline combines RMDetect with the RNA 3D Motif Atlas and the Rfam database in order to automate the building process of Bayesian network.[38] On a large-scale test, RMDetect extracted more than 22,000 modules in all PDB files and, identified 977 internal loops and 17 hairpin modules with clear discriminatory power. JAR3D is another program for identifying modules from sequences.[39] It uses hybrid stochastic context-free grammars technique to model the nested base pairs and insertions, and uses Markov random fields to handle base triples. JAR3D assigns acceptance/rejection thresholds for motif groups and reduces the false positive rate, which is a central challenge in matching novel sequences to motifs. The comparison of JAR3D with RMDetect showed the same output on 257 sequences except 63 sequences. JAR3D was designed to incorporate automatically new motifs as they are solved and accumulate in the database. The modules identification programs have been shown to be able to improve both secondary structure and tertiary structure predictions.[36,40]

**Electrostatic interactions in RNA structures**

RNA structure modeling has a particular difficulty to overcome – the strong electrostatic interaction between nucleic acids.[7-8] Since RNA chains are highly negatively charged, they will not fold unless the negative charges are neutralized by metal ions. The roles of metal ions are not only limited to charge neutralization, but also include binding to specific locations to stabilize the structure, as well as directly mediating catalysis in some ribozymes. Previous knowledge-based potential functions, such as RASP,[25] Rosetta,[28-29] KB,[30] and 3dRNAscore[33] energy functions, implicitly model the electrostatic interactions by inferring energies from observed frequencies of geometrical features. However, such implicit treatments cannot predict the dependence of RNA properties on such as salt concentrations, multivalent ions, or global versus local binding of ions.

The counterion condensation (CC) and the Poisson–Boltzmann (PB) theory have been developed to model the electrostatic interactions. However, the CC theory only works well at very dilute ion concentrations while not at finite concentrations,[41] and the PB theory ignores the correlation between ions and significantly underestimates the effect of multivalent ions in stabilizing RNAs.[42-43]

Recently, the tightly bound ion (TBI) model has been developed by accounting for fluctuations and ion-ion correlations.[44-45] The TBI model separates the tightly bound ions from the diffusive ions in solution, and explicitly accounts for the correlation between the tightly bound ions and the discrete binding modes. The electrostatic free energy is calculated from the partition function Z, which is a summation over the partition functions of all possible binding models. To enumerate the discrete ion-binding modes, the tightly bound region near the RNA surface is divided into $N$ cells and all the possible positions of the tightly bound ions among the cells are then enumerated. The diffusive ions are treated with the mean-field PB theory. The TBI model was originally based on CG DNA helices and later extended to treat RNA tertiary folds at the all-atoms level.[46] The TBI model has been shown to improve the predictions of effects of multivalence ions on RNA helices, hairpins, pseudo-knots and, the stability of RNA kissing complexes over wide ranges of ion concentrations when combined with a GC model.[44-49] However, the TBI model is not perfect when it concerns the ion binding in the vicinity of nucleic acids surface. Nevertheless, even all-atom force fields, such as AMBER, do not see the exact electrostatic interaction by using the point-charge approximation.[50]

To develop an accurate and yet efficient energy function for the electrostatic interactions, machine-learning methods may be helpful. For example, Li et al. optimized the parameters in the AMOEBA polarizable force field using ML, genetic algorithm techniques and ab initio data from quantum mechanics (QM) calculations. [51] Their work showed that the ML can be used in the parameterization step of traditional force fields and achieve better performance. Bereau et al. developed a many-body non-additive potential for small neutral organic and biologically relevant molecules.[52] They modeled the intermolecular interactions with sophisticated physical models, e.g., multipole rather than point-charge electrostatics, non-additive rather than pairwise additive polarization, and relied on machine learning to optimize the parameters. The model allows accurate calculations of electrostatics, charge penetration, polarization, repulsion, and many-body dispersion. [52]

The force field developed by Wang et al. for water molecules based on neural network (NN) is particularly interesting.[53] The force field uses the many-body expansion up to binary interactions and NN representation of atomic energies, employing an electrostatic embedding scheme. For each one-body or two-body term, only the water molecules in the QM regions are treated with NN representations, while those in the MM regions serve as background charges to provide electrostatic embedding. This is similar to the separation of tightly bound ions from diffusive ions found in the TBI model. [44-45] The authors built two sets of NN based force fields: nonpolarizable and polarizable force fields, and showed that the first one has already behaved well, since the polarization and many-body effects were implicitly considered in the electrostatic embedding scheme. Furthermore, according to the authors, the force field shows high level of QM accuracy but low computational costs. Presumably, similar ideas may be implemented to treat the strong electrostatic effect in RNA molecules.

In a recent review, Popelier discussed the next-next generation force field, called QCTFF (Quantum Chemical Topology Force Field).[54] The author suggested that only a machine-learning model could cope with the complexity of the atomic environment and learn how energy quantities vary with the coordinates of the atomic neighbors. QCTFF is a machine learning method based on kriging. The term kriging refers to a group of statistical techniques that interpolate the value of a random field at an unobserved location from the observation of its values at nearby locations. The kriging algorithm maximizes the likelihood function of recovering the observed input data, which come from pure QC computations or experimental data. The model predicts the monopole and dipole moments, self-energy, and exchange energy as function of the nuclear coordinates of the atoms. The force field abolishes all traditional force field expressions such as the Hooke's law and the Lennard-Jones potential model. The force field has been tested on several small molecules such as water, methanol, propane, and N-methylacetamide and gave very accurate results.[54]

**Learning from protein structure predictions**

RNA structure prediction may borrow ideas from its sister field, the protein structure prediction, where researchers have developed many ingenious machine learning algorithms for predicting protein structures.[55-57] Therefore, it is interesting to check the most recent advances in this field and see if they can be applied in RNAs.

In the recent Critical Assessment of Protein Structure Prediction (CASP13),[58] a blind assessment of the state of the protein folding predictions, the AlphaFold system predicted high-accuracy structures for 24 out of 43 free-modeling domains, showing significantly better performance than the second-best method.[59] It demonstrated the power of machine learning method in protein structure predictions.

AlphaFold created three neural networks for free-modeling predictions. The first network, maybe the most important one, utilizes evolutionary covariation data to predict the distances between pairs of residues. The network consists of 220 two-dimensional residual blocks with 128 channels and dilated 3 × 3 convolutions, as well as dropout and batch normalization layers. A distance potential is also inferred from the negative log-likelihood of the distances predicted by the network. The second neural network, called GDT-net, takes the predicted distances, along with the multiple sequence alignment (MSA) features, the $C_\beta$ coordinates and sine/cosine of the torsion angles as inputs, and then predicts the GDT_TS score of the given candidate structure. The GDT-net begins with a deep resnet stack, followed by 18 residual blocks with 3 × 3 dilated convolutions, a mean pooling layer, and a softmax layer. The third neural network uses an end-to-end trained generative model of backbone torsion angles, conditioned on the sequenced and MSA features to create libraries of fragments. The major components of the networks include a 2D residual network and a 1D convolutional LSTM encoder. The first component encodes the conditioning data and the second models the sequence dependence of the torsional angles. The authors tried three different combinations of the above components with simulated annealing algorithm and direct gradient descent optimization of the potential for structure prediction. It was found that the three systems performed similarly. The authors drew two main conclusions. First, the methods relied heavily on distance predictions based on coevolutionary data; second, the good results were due to the fact that deep learning allows extract features from data without making heuristic assumptions about the data.

The coevolutionary data have been also used in RNA structure prediction and significantly improved its performance.[60-62] It has been shown by Xiao's group that the incorporation of the contact information obtained by direct coupling analysis (DCA) of nucleotide coevolution greatly increased the accuracy of their 3dRNA system, particularly in predicting multi-branch junctions.[62-65] The DCA analysis in the above works assumed a global statistical model of nucleotide correlation, such as a generalized Potts model. However, machine-learning techniques can further improve the performance, since it makes no assumption on the data and can predict distance rather than just contacts. For example, He et al. compared the performance of different approaches of inferring RNA contacts and found that a deep learning model of fully convolutional neural network improved the performance of DCA.[66] More examples can be found in the literature.[67-68]

It should be noted that a simple transferring of the approaches developed for protein structure prediction may not work for RNAs, since the latter is very flexible in the structures,[69] and the available experimental structures are far less than the former. As of the year 2020, the number of PDB-only structures deposited in the PDB database has been more than 152,000; whereas the number of RNA-only was less than 1,500. According to the latest version (14.2) of the Rfam database,[70] there are 3024 reported RNA families while only 99 families have 3D structural information. Therefore, more sophisticated machine learning techniques, for example few-shot learning or Meta-learning approaches,[71-73] need to be considered.

**Conclusions and Perspectives**

The adoption of machine learning approaches in RNA structure predictions has led to many promising results, opening a new way of thinking and solving the problem. Unlike traditional approaches of developing knowledge-based potentials, multilayer perceptrons need no explicit form of functions as input and they are very flexible in the choice of input features. Furthermore, the convolutional neural networks are able to discover relevant features on their own, if they are provided with a proper dataset and well trained. These characteristics may help to overcome difficult problems such as modeling many-body interactions, for which actual function form cannot be easily determined.

There are several challenges with regards to this research direction. First, the small number of experimental RNA structures limits the performance of machine learning-based approaches. To alleviate the problem, few-shot learning and

Meta-learning techniques may be considered,[71-73] and physical knowledge could be incorporated into the neural networks as *a prior* to reduce the hypothesis space or direct the optimization process. Second, neural networks are usually treated as inscrutable black box, lacking transparency and explanations. It is necessary to collaborate with researches in machine learning field to design new network architectures with higher interpretability or new analyzing tools to dissect the network. In summary, the research on utilizing machine learning approaches for RNA structure predictions is undoubtedly still premature, and more effort needs to be made to push forward the front of the field.

**Acknowledgements:**

The authors acknowledge the financial support from National Natural Science Foundation of China, the computational support from High Performance Computing Center in the Collaborative Innovation Center of Advanced Microstructures, Nanjing University. The authors also acknowledge the editing service provided by Enago (www.engao.cn).

**References**

[1] Mercer T R, Dinger M E, and Mattick J S. 2009 Nat. Rev. Genetics 10 155

[2] Geisler S, and Coller J 2013 Nat. Rev. Mol. Cell Biol.14 699

[3] Cech T R, and Steitz J A 2014 Cell 157 77

[4] Morris K V, and Mattick J S 2014 Nat. Rev. Genetics 15 423

[5] Anastasiadou E, Jacob L S, and Slack F J. 2018 Nat. Rev. Cancer 18 5

[6] Miao Z, Adamiak R W, Antczak M, et al. 2017 RNA 23 655

[7] Chen S J 2008 Annu. Rev. Biophys. 37 197

[8] Sun L Z, Zhang D, and Chen S J 2017 Ann. Rev. Biophys. 46 227

[9] Sponer J, Bussi G, Krepl M, Banas P, Bottaro S, Cunha R A, Gil‐Ley A, Pinamonti G, Poblete S, Jurecka P, Walter N G, and Otyepka M 2018 Chem. Rev. 118 4177

[10] Dans P D, Gallego D, Balaceanu A, Darre L, Gomez H, and Orozco M 2019 Chem 5 51

[11] Shi Y Z, Wu Y Y, Wang F H, Tan Z J 2014 Chin. Phys. B 23 078701

[12] Goodfellow I, Bengio Y, and Courville A. 2016 Deep learning. Adaptive computation and machine learning (Cambridge, Massachusetts: The MIT Press) pp. 197-200

[13] Silver D, Huang A, Maddison C J, Guez A, Sifre L, van den Driessche G, Schrittwieser J, Antonoglou1 I, Panneershelvam V, Lanctot M, Dieleman S, Grewe D, Nham H, Kalchbrenner N, Sutskever I, Lillicrap T, Leach M, Kavukcuoglu K, Graepel T, and Hassabis D 2016 Nature 529 484

[14] Alipanahi B, Delong A, Weirauch M T, and Frey B J 2015 Nat. Biotech. 33 831

[15] Zhou J, and Troyanskaya O G 2015 Nat. Methods 12 931

[16] Carleo G, and Troyer M. 2017 Science 355 602

[17] Carrasquilla J, and Melko R G. 2017 Nature Physics 13 431

[18] Yonemotoa H, Asai K, and Hamada M 2015 Comput. Biol. and Chem. 57 72

[19] Ray S S and Pal S K 2013 IEEE/ACM Trans. on Compt. Biol. and Bioinformatics 10 1

[20] Koessler D R, Knisley D, Knisley J, and Haynes T 2010 BMC Bioinformatics 11 S21

[21] Tan Y L, Feng C J, Jin L, Shi Y Z, Zhang W B, and Tan Z J 2019 RNA 25 793

[22] Yang Y, Gu Q, Zhang B G, Shi Y Z, and Shao Z G 2018 Chin. Phys. B 27 038701

[23] Wang Y Z, Li J, Zhang S, Huang B, Yao G, and Zhang J 2019 Molecular Biol. 53 118

[24] Tsai J, Bonneau R, Morozov A V, Kuhlman B, Rohl C A, Baker D 2003 Proteins 53 76

[25] Capriotti E, Norambuena T, Marti-Renom M A, and Melo F. 2011 Bioinformatics 27 1086

[26] Li J, Zhu W, Wang J, Li W F, Gong S, Zhang J, and Wang W 2018 Plos Comput. Biol. 14 e1006514

[27] Simonyan K, and Zisserman A 2014 arXiv:1409.1556v6

[28] Das R, Baker D. 2007 Proc. Natl. Acad. Sci. USA 104 14664

[29] Das R, Karanicolas J, and Baker D. 2010 Proc. Natl. Acad. Sci. USA 7 291

[30] Bernauer J, Huang X H, Sim A, and Levitt M. 2011 RNA 17 1066

[31] Cruz J A, Blanchet M F, Boniecki M et al. 2012 RNA 18 610

[32] Miao Z, Adamiak R W, Blanchet M F et al. 2015 RNA 21 1066

[33] Wang J, Zhao Y J, Zhu C Y, and Xiao Y 2015 Nuc. Acids Res. 43 e63

[34] Frellsen J, Moltke I, Thiim M, Mardia K V, Ferkinghoff-Borg J, Hamelryck T 2009 Plos

Comput. Biol. 5 e1000406

[35] Wang Z, Xu J 2011 Bioinformatics 27 i102

[36] Miao Z, Westhof E 2017 Annu. Rev. Biophys. 46 483

[37] Cruz J A, Westhof E 2011 Nature Methods 8 513

[38] Theis C, Siederdissen C H, Hofacke I L, Gorodki J 2013 Nuc. Acids Res. 41 9999

[39] Zirbel C, Roll J, Sweeney B A, Petrov A I, Pirrung M, Leontis N B 2015 Nuc. Acids Res. 43 7504

[40] Theis C, Zirbel C L, Siederdissen C H, Anthon C, Hofacker I L, Nielsen H, Gorodkin J 2015 PLOS ONE 10 e0139900

[41] Manning G S 2007 J. Phys. Chem. B. 111 8554

[42] Baker N A 2005 Curr. Opin. Struct. Biol. 15 137

[43] Xiong G, Xi K, Zhang X, and Tan Z J 2018 Chin. Phys. B 27 018203

[44] Tan Z J, and Chen S J 2005 J. Chem. Phys. 122 44903

[45] Tan Z J, and Chen S J 2006 Biophys. J. 90 1175

[46] Tan Z J, and Chen S J 2010 Biophys. J. 99 1565

[47] Tan Z J, and Chen S J 2011 Biophys. J. 101 176

[48] Shi Y Z, Jin L, Feng C J, Tan Y L, and Tan Z J 2018 Plos Comput. Biol. 14 e1006222

[49] Jin L, Tan Y L, Wu Y, Wang X, Shi Y Z, and Tan Z J 2019 RNA 25 1532

[50] Wang J M, Cieplak P, Li J, Wang J, Cai Q, Hsieh M J, Lei H X, Luo R, and Duan Y 2011 J. Phys. Chem. B 115 3100

[51] Li Y, Li H, Pickard F C, Narayanan B, Sen F G, Chan M, Sankaranarayanan S, Brooks B R, Roux B 2017 J. Chem. Theory Comput. 13 4492

[52] Bereau T, DiStasio R A, Tkatchenko A, and Lilienfeld O A 2018 J. Chem. Phys. 148 241706

[53] Wang H, Yang W 2018 J. Phys. Chem. Lett. 9 3232

[54] Popelier P L A 2016 Physica Scripta 91 033007

[55] Hanson J, Paliwal K, Litfin T, Yang Y, and Zhou Y Q 2018 Bioinformatics 34 4039

[56] Wang S, Sun S, and Xu J B 2017 Proteins 86 67

[57] Kandathil S M, Greener J G, and Jones D T 2019 Proteins 87 1179

[58] Kryshtafovych A, Schwede T, Topf M, Fidelis K, and Moult J 2019 Proteins 87 1011

[59] Senior A W, Evans R, Jumper J, Kirkpatrick J, Sifre L, Green T, Qin C, Zidek A, Nelson A,

Bridgland A, Penedones H, Petersen S, Simonyan K, Crossan S, Kohli P, Jones D T, Silver D, Kavukcuoglu K, and Hassabis D 2019 Nature 577 706

[60] Weinreb C, Riesselman A J, Ingraham J B, Gross T, Sander C, Marks D S 2016 Cell 165 963

[61] Leonardis E D, Lutz B, Ratz S, Cocco S, Monasson R, Schug A, and Weigt M 2015 Nuc. Acid Res. 43 10444

[62] Wang J, Mao K, Zhao Y J, Zeng C, Xiang J, Zhang Y, and Xiao Y 2017 Nuc. Acids, Res. 45 6299

[63] Zhao Y, Huang Y, Gong Z, Wang Y, Man J, and Xiao Y 2012 Scientific Reports 2 734

[64] Wang J and Xiao Y 2017 Current Protocols in bioinformatics 57 5.9.1

[65] Wang J, Wang J, Huang Y, and Xiao Y 2019 Intern. J . Mol. Sci. 20 4116

[66] He X L, Li S M, Ou X J, Wang J, and Xiao Y 2019 Comm. in inform. and syst. 19 279

[67] Singh J, Hanson J, Paliwal K, and Zhou Y Q 2019 Nature Comm. 10 5407

[68] Zhang H, Zhang Q, Ju F, Zhu J, Gao Y, Xie Z, Deng M, Sun S, Zheng W M, and Bu D B 2019 BMC Bioinformatics 20 537

[69] Bao L, Zhang X, Jin L, and Tan Z J 2016 Chin. Phys. B 25 018703

[70] Kalvari I, Argasinska J, Quinones-Olvera N, Nawrocki E P, Rivas E, Eddy S R, Bateman A, Finn R D, Petrov A 2018 Nuc. Acids, Res. 46 D335

[71] Wang J X, Nelson Z K, Tirumala D, Soyer H, Leibo J Z, Munos R, Blundell C, Kumaran D, and Botvinick M 2017 arXiv:1611.05763v3

[72] Zhou Z H 2018 National Science Review 5 44

[73] Wang Y, Yao Q, Kwok J K, and Ni L M 2020 ACM Computing Surveys 53 63